\PassOptionsToPackage{dvipsnames}{xcolor}
\documentclass[10pt,sigconf,nonacm,screen]{acmart}

\usepackage[all]{nowidow}
\usepackage{xcolor}
\usepackage{subcaption}
\usepackage{hyperref}
\usepackage{acronym}
\usepackage{tikz}
\usepackage[capitalise,noabbrev]{cleveref}
\crefformat{section}{\S#2#1#3} 
\crefformat{subsection}{\S#2#1#3}
\crefformat{subsubsection}{\S#2#1#3}
\crefrangeformat{section}{\S#3#1#4 to~\S#5#2#6}
\crefrangeformat{subsection}{\S#3#1#4 to~\S#5#2#6}
\crefrangeformat{subsubsection}{\S#3#1#4 to~\S#5#2#6}

\usepackage{todonotes}
\definecolor{darkblue}{rgb}{0.0, 0.0, 0.55}
\begin{document}

\author{Tim C. Rese}
\affiliation{%
    \institution{TU Berlin}
    \city{Berlin}
    \country{Germany}}
\email{tr@3s.tu-berlin.de}
\orcid{0009-0008-0185-8339}

\author{David Bermbach}
\affiliation{%
    \institution{TU Berlin}
    \city{Berlin}
    \country{Germany}}
\email{db@3s.tu-berlin.de}
\orcid{0000-0002-7524-3256}

\title{Poster: Towards an Application-Centric Benchmark Suite for Spatiotemporal Database Systems}



\begin{abstract}
    Spatiotemporal data play a key role for mobility-based applications and are their produced volume is growing continuously, among others, due to the increased availability of IoT devices. 
    When working with spatiotemporal data, developers rely on spatiotemporal database systems such as PostGIS or MobilityDB.
		For better understanding their quality of service behavior and then choosing the best system, benchmarking is the go-to approach.
		Unfortunately, existing work in this field studies only small isolated aspects and a comprehensive application-centric benchmark suite is still missing.
		
		In this paper, we argue that an application-centric benchmark suite for spatiotemporal database systems is urgently needed.
		We identify requirements for such a benchmark suite, discuss domain-specific challenges, and sketch-out the architecture of a modular benchmarking suite.
\end{abstract}

\maketitle
\begin{tikzpicture}[remember picture,overlay]
    \node[anchor=south west, xshift=1.8cm, yshift=0.5cm] at (current page.south west) {
      \begin{minipage}[t]{0.45\textwidth}
      \footnotesize
      © 2025 IEEE. Personal use of this material is permitted. Permission from
      IEEE must be obtained for all other uses, in any current or future media,
      including reprinting/republishing this material for advertising or promotional
      purposes, creating new collective works, for resale or redistribution to servers
      or lists, or reuse of any copyrighted component of this work in other works.\\
      DOI: 10.1109/IC2E65552.2025.00016
      \end{minipage}
    };
\end{tikzpicture}
\section{Introduction}
\label{sec:introduction}

Spatiotemporal data play an evergrowing role in modern cloud applications, ranging from routing car traffic to analyzing the spread of diseases ~\cite{waller1997hierarchical}.
Moving object data (MOD) is a subset of such data that describes the movement of one or multiple objects over time, which can be represented as a trajectory through multidimensional space or as a sequence of points in time. 
Depending on the application, MOD can heavily vary in characteristics, e.g., the number of tracked objects or average trajectory length ~\cite{rese2025evaluating}.

Current database systems focused on spatiotemporal or moving object data need to be able to handle these varying datasets efficiently.
Additionally, user interaction patterns with such data may also differ, e.g., the frequency of writes and reads, the complexity of queries, and the required parallelism of operations. 
While benchmarks for moving object data exist, e.g., BerlinMOD which covers parts of the benchmarking process, a comprehensive benchmarking suite that includes a start-to-end benchmarking tool is still missing ~\cite{zimanyi2020mobilitydb}.
In this paper, we propose a new benchmarking suite for spatiotemporal and moving object database systems, which includes the entire benchmarking process within its scope. 
We consider client interaction patterns, dataset characteristics, query types, and database systems within our vision. 
\section{Background and Related Work}
\label{sec:background}
Spatiotemporal data consist of objects' spatial and temporal information, usually referring to their latitude/longitude position and an associated position. 
Moving object data (MOD) refer to tracking such objects' position over time, e.g., tracking a car through city traffic or a plane along its path. 

MOD database systems usually provide unique features, thus, optimizing for high performance access. 
Previous work on benchmarking MOD database systems so far covers only parts of the benchmarking process, e.g., data and query generation.
BerlinMOD is a benchmark for moving object data that includes its own data generator for car data and includes a wide range of queries, but does not include a client interaction model or a database system to run the benchmark against~\cite{duntgen2009berlinmod}.
Its dataset and queries have been included in a variety of system design papers, such as MobilityDB and SECONDO, two database systems that provide specific support for MOD~\cite{zimanyi2020mobilitydb,guting2005secondo}.
Other work such as GMOBench focuses on tracking an object as it changes its transportation mode, e.g., from walking to driving, and includes a data generator and a query generator, but does not include a client interaction model or support for more than one database system~\cite{xu2015gmobench}.
Other work specifically focuses on evaluating moving object indexes, however it does this outside of an actual database system and only with synthetically generated datasets based on realistic road networks ~\cite{chen2008benchmark}. 
Thus, it can explore the effects of index choice but cannot be used to study the end-to-end performance of a database system.

\section{Benchmark Suite Requirements}
\label{sec:requirements}
\begin{figure}[ht]
    \centering
    \includegraphics[width=0.5\textwidth]{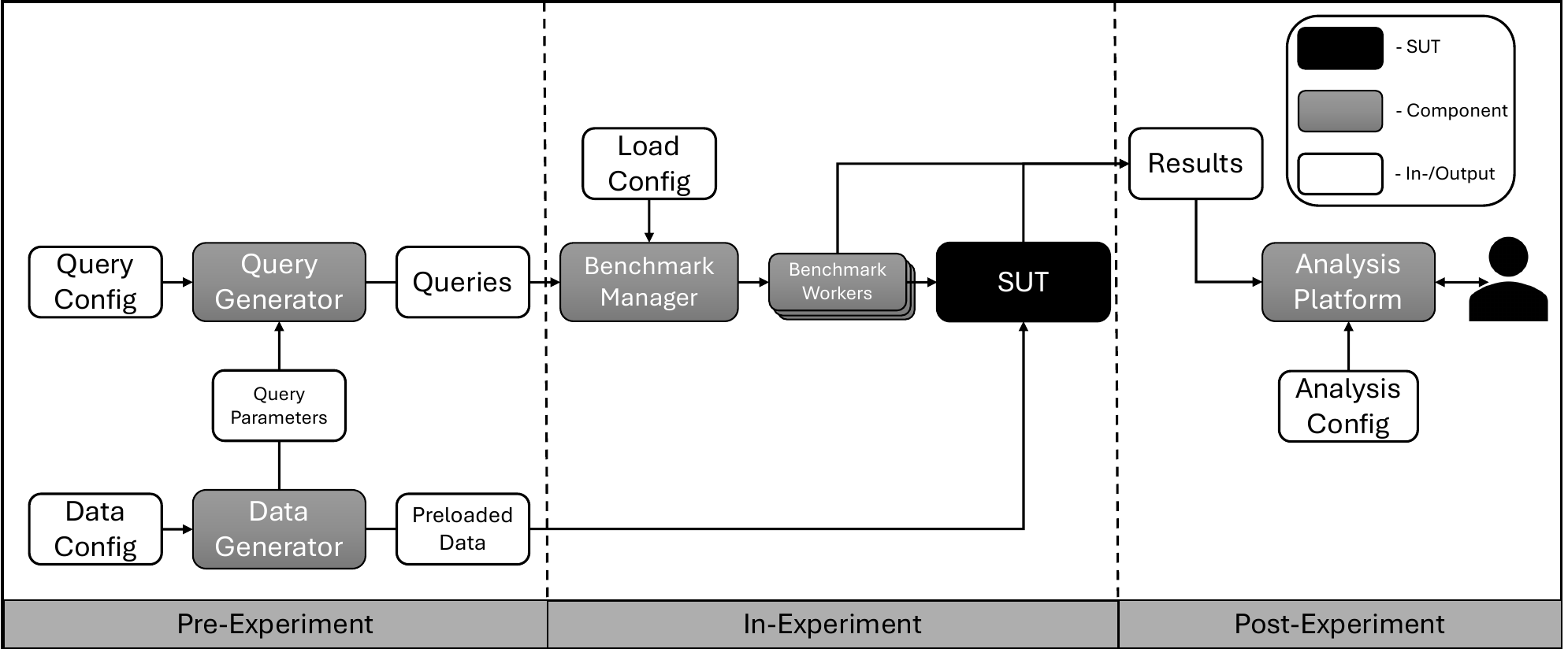}
    \caption{Each component is designed as a separate module with unique configuration options, which allows users to insert their own implementation. The analysis platform should be able to evaluate multiple experiments at once.}
    \label{fig:architecture}
\end{figure}

Building an application-centric spatiotemporal benchmarking suite that covers all aspects of benchmarking sets specific requirements that need to be addressed that extend past previously defined database benchmarking requirements ~\cite{bermbach2017benchfoundry}:
\paragraph*{\textbf{Flexible Workload (R1)}}
As clients usually interact with an application in parallel, evaluating database system performance with a varying number of workers is essential. 
The load generation coordinator needs to trigger a configurable number of workers to generate load, as well as be able to adjust query complexity and vary the balance of read/write requests. 

Additionally, query parameters and the amount of queried data can vary depending on the application. 
Therefore, the suite should allow users to specify their application characteristics, including data scale, query complexity/parameters, and the number of benchmarking workers that generate the load for the System Under Test (SUT).
\paragraph*{\textbf{Platform Extensibility (R2)}}
As the field of spatiotemporal and moving object database systems is varied, the suite should make as little assumptions as possible about the SUT.
Database systems tend to differ in their query dialect to a varying degree, providing a query generation tool that converts a provided configuration file into different languages is required to simplify suite use.
Optimally, the user defines a query using a query template, which is then translated into different query languages automatically. 
This tool should be extensible to support further query languages, as well as allow users to insert their own query generation logic if required.

Additionally, certain systems may only require a single node to perform adequately, while others rely on a distributed setup to handle the data size and worker requests. Therefore, both single-and distributed benchmark configurations should be supported by such a suite.
\paragraph*{\textbf{Support of Independent Analysis (R3)}}
Depending on the application, different metrics may be key. Therefore, the suite should provide support to customize the analysis configuration, which would then tune the analyzed metrics, such as average query time/type, throughput, resource usage, index build time, and more. 

\section{Possible Design of Benchmark Suite}
\label{sec:approach}

We propose a comprehensive benchmarking suite for spatiotemporal and moving object database systems that supports the entire benchmarking process.
We envision components for such a system, which aim to cover the requirements outlined in \cref{sec:requirements}, and the design of which is shown in \cref{fig:architecture}.

We picture the benchmarking suite to be split into three phases: \textit{Pre-Experiment}, \textit{In-Experiment}, and \textit{Post-Experiment}.

During the \textit{Pre-Experiment} phase, several components are responsible for preparing the experiment. 
The data generator is given configuration files, which define the application and data characteristics, as well as the data scale. 
The generated data are then stored in the database system. 

Said database system would be deployed in a single-node or distributed setup depending on the configuration, which would partially satisfy \textbf{R2}.
Data are then indexed accordingly.

In our vision, a load generation coordinator is responsible for organizing the load generation process, which would include one or several distributed workers running queries against the database system. 
The coordinator is responsible for distributing tasks among available workers, could however also run a single worker, which would fulfill parts of \textbf{R1}. 

The coordinator receives the queries to run from a query generator, which creates queries based on a user-provided query configuration file.
The user should be able to fine tune data scale and query complexity using the file, which would satisfy the remainder of \textbf{R1}.
Additionally, the query generator would be communicating with the data generator to ensure query compatibility with the generated data.
The query generator would also be responsible for translating queries into the database system's query language, which fulfills \textbf{R2}.

Once load has been distributed among workers, the \textit{In-Experiment} phase begins. Depending on the configuration, experiments can now be conducted, with the database system providing query results and resource usage metrics.

In the \textit{Post-Experiment} phase, the analysis platform collects the results from the database system and the workers, and analyzes the results based on a user-provided analysis configuration file, fulfilling \textbf{R3}.
The user should be able to interact with the platform to perform custom analyses, such as visualizing the results.

\section{Conclusion}
\label{sec:conclusion}
In this paper, we highlighted how current spatiotemporal benchmarking tools include only parts of the benchmarking process. 
We detailed our vision for a benchmarking suite for spatiotemporal database systems that provides a start-to-end benchmarking pipeline that enables users to fine tune their desired benchmarking process.
In our vision, we covered how we picture the suite to be a modular component system that allows for extension and customization to best adapt to user needs, while highlighting how we would ensure that the suite is compatible with various applications and database systems.


\balance

\bibliographystyle{ACM-Reference-Format}
\bibliography{bibliography.bib}

\end{document}